# Ranking Robustness Under Adversarial Document Manipulations

Gregory Goren
gregory.goren@campus.techion.ac.il
Technion — Israel Institute of Technology

Oren Kurland
kurland@ie.technion.ac.il
Technion — Israel Institute of Technology

Moshe Tennenholtz
moshet@ie.technion.ac.il
Technion — Israel Institute of Technology

Fiana Raiber
fiana@oath.com
Yahoo Research

## ABSTRACT

For many queries in the Web retrieval setting there is an on-going ranking competition: authors manipulate their documents so as to promote them in rankings. Such competitions can have unwarranted effects not only in terms of retrieval effectiveness, but also in terms of ranking *robustness*. A case in point, rankings can (rapidly) change due to small indiscernible perturbations of documents. While there has been a recent growing interest in analyzing the robustness of classifiers to adversarial manipulations, there has not yet been a study of the robustness of relevance-ranking functions. We address this challenge by formally analyzing different definitions and aspects of the robustness of learning-to-rank-based ranking functions. For example, we formally show that increased regularization of linear ranking functions increases ranking robustness. This finding leads us to conjecture that decreased variance of any ranking function results in increased robustness. We propose several measures for quantifying ranking robustness and use them to analyze ranking competitions between documents' authors. The empirical findings support our formal analysis and conjecture for both RankSVM and LambdaMART.

## 1 INTRODUCTION

In adversarial retrieval settings (e.g., the Web) there is an on-going ranking competition for many queries: authors of some Web documents manipulate them so as to have them ranked high. The ranking competition can have various undesirable effects. First, ranking effectiveness can degrade due to adversarial changes of documents that result in having them ranked higher than they should; i.e., black-hat search engine optimization (SEO) [11]. Furthermore, rankings can potentially (rapidly) change due to small document perturbations that might be indiscernible.

Motivated by the ranking competitions that take place in the adversarial Web retrieval setting, and the growing body of work on robust classification — specifically, with respect to adversarial manipulations [4, 6–10, 14, 20, 21, 25], we present the first (to the best of our knowledge) theoretical and empirical study of the robustness of document relevance-ranking functions to document manipulations. Our focus is learning-to-rank-based functions [13] where a document-query pair is represented as a feature vector.

We start by adapting a basic classifier-robustness notion used in recent work on robust classification [7, 20] to the case of a document ranking function. We formally analyze implications of applying this notion and highlight a notable drawback: the treatment of documents independently of each other — i.e., this is a pointwise robustness perspective. However, ranking depends on the relative retrieval scores of documents. Hence, we formulate a definition of pairwise robustness that addresses the effect of small document changes on the relative ranking of pairs of documents. We formally analyze the implications of applying our pairwise robustness definition and the connections with pointwise robustness.

The different definitions of robustness that we propose are based on a worst-case scenario; namely, quantifying the minimal document change needed to change a ranking. Using the definitions to compare the robustness of different ranking functions can be quite difficult. Thus, we explore an additional aspect of robustness which we term *stability* (cf. [20]): changes of retrieval scores and relative ranking with respect to a *given* fixed change of a document. We then establish formal connections for linear (in features) ranking functions between the extent of their regularization and their stability. Motivated by these formal findings, we state a *variance conjecture*: the higher the variance of a learned ranking function, the less robust the rankings it induces. A ranking is considered robust if it does not significantly change due to small documents' changes. We propose a few methods of measuring ranking robustness.

While our motivation is to address documents' changes introduced by incentivized authors, our formal analysis makes no assumptions on the cause and nature of documents' changes. Thus, the analysis constitutes a general treatment of ranking robustness under document manipulations. Since, in practice, documents' changes in competitive retrieval settings are often incentivized (adversarial), we use for evaluation a recently published dataset of document ranking competitions held between students [17]. The mere motivation to change documents in the competitions was to have them ranked high as possible.

The analysis of the ranking competitions provides support to the formal analysis and the variance conjecture. First, increased regularization of a linear ranking function (RankSVM [12]), which leads to reduced variance, results in improved ranking robustness. Second, increased regularization of a non-linear ranking function, namely, the state-of-the-art LambdaMART method [27], also results in improved ranking robustness. Third, LambdaMART induces rankings that are less robust than those induced by RankSVM; the former has higher variance than the latter.

Our contributions can be summarized as follows:

- We present the first formal and empirical analysis of the robustness of learning-to-rank-based relevance ranking functions to (adversarial) manipulations of documents.
- We formally demonstrate, and provide empirical support to, the connection between regularization of linear learning-to-rank functions and ranking robustness.
- Motivated by our formal findings, we post a conjecture about the connection between the variance of a ranking function and ranking robustness, and provide empirical support.

## 2 RANKING ROBUSTNESS

We assume that the following have been fixed: a query $q$, a document corpus $\mathcal{D}$, and a document relevance ranking function $f$. As is the case for many learning-to-rank-based retrieval methods [13], the ranking function takes as input a feature vector $\vec{d} \in \mathfrak{R}^m$ that represents the pair of document $d$ ($\in \mathcal{D}$) and the query[1] $q$; $\vec{d}[i]$ is the i'$th$ component of $\vec{d}$. Features can model query-document relations, query-independent properties of documents and document-independent properties of the query [13]. The output of the ranking function (wlog) is a retrieval score $f(\vec{d}) \in \mathfrak{R}_+$ used to induce a ranking over the corpus. We assume that ties of retrieval scores are consistently broken (e.g., using documents' IDs). We use $L_2$ norms of vectors. For a linear (in features) ranking function $f$, $f(\vec{d}) \stackrel{def}{=} \vec{w}\vec{d} + b$; $\vec{w}$ is the weight vector and $b$ is the intercept.

Our goal is to define and quantify notions of the robustness of ranking functions $f$ to small perturbations of documents. Rather than directly address documents' changes (e.g., with respect to content, hyperlink, hypertext, etc.), we study the effects of the resultant perturbations of the corresponding feature vectors.

We first motivate our analysis of ranking robustness in Section 2.1. Then, in Section 2.2 we adapt a recently used classifier-robustness definition [6, 20] to ranking functions. The definition addresses documents independently. Since ranking is determined by the relative retrieval scores of documents, we study the notion of pairwise document robustness in Section 2.3; that is, we examine effects of documents' perturbations on the relative ranking of two documents. Given our formal findings in Section 2.3, we post a conjecture about the connection between ranking robustness and the variance of a ranking function in Section 2.4. Finally, in Section 2.5 we present a few methods to measure ranking robustness.

### 2.1 Motivation

Much of the practical motivation for the recent line of work on robustness of classifiers comes from the vision realm [7, 8, 10, 20, 25]. The assumption is that small perturbations of images not discernible by humans should not result in changes to classification decisions.

We make a similar assumption about the ranking of documents. That is, a ranking should not (significantly) change as a result of small, potentially indiscernible, perturbations of documents[2]. A case in point, users might suspect the validity of rankings given rapid changes that are hard to explain (a.k.a., "explainable IR").

To further support our premise, we appeal to the cluster hypothesis which states that "closely associated documents tend to be relevant to the same requests" [22]. An important operational manifestation of the hypothesis is the premise that "similar documents should receive similar retrieval scores" [5]. That is, by the cluster hypothesis, similar documents would be relevant to the same requests (i.e., queries). Therefore, their retrieval scores which reflect relevance status should not be very different. In other words, small perturbations of documents should not result in significant changes of retrieval scores, and hence, significant ranking changes[3].

Accordingly, below we focus on the effects of documents' perturbations on induced rankings; i.e., ranking robustness. The formal treatment of the effects on document relevance, and consequently ranking effectiveness, is an intriguing research venue at its own right which we leave for future work.

### 2.2 Pointwise Robustness

The recently adopted notion of classifier robustness was defined based on the minimal change of an object required to change the class to which the object is classified [7, 20]. We conceptually adapt this robustness notion to the case of retrieval scores. We term this robustness "pointwise" as documents are considered independently — i.e., the effects of the change of a document retrieval score on its relative ranking with respect to other documents are not considered. To simplify the following definitions and analysis, and without loss of generality, we focus, unless otherwise stated, on the increase of a retrieval score rather than a change which could be negative.

DEFINITION 1 (POINTWISE ROBUSTNESS). *The pointwise robustness of ranking function $f$ with respect to document $d$ and query $q$ is:* $\rho_{point}(f; \vec{d}, q) \stackrel{def}{=} \min_{\vec{v} \in \mathfrak{R}^m} ||\vec{v}||$ *s.t.* $f(\vec{d} + \vec{v}) > f(\vec{d})$. *The robustness with respect to $q$ is the expectation over all documents $d$ ($\in \mathcal{D}$) sampled using some distribution $\mathbb{P}(\mathcal{D})$:* $\rho_{point}(f; q) \stackrel{def}{=} \mathbb{E}_{d \sim \mathbb{P}(\mathcal{D})} \rho_{point}(f; \vec{d}, q)$.

In other words, the pointwise robustness with respect to document $d$ is the minimal extent of a change (i.e., norm of a vector) needed to be applied to $\vec{d}$ so as to increase $d$'s retrieval score.

We now analyze the pointwise robustness of linear ranking functions: $f(\vec{d}) \stackrel{def}{=} \vec{w}\vec{d} + b$. RankSVM [12] and coordinate ascent [15] are examples of commonly used, effective linear ranking functions.

---

[1] We omit $q$ from the feature-vector notation as it is fixed throughout the section.

[2] In contrast to spamming, white-hat SEO attempts to promote documents in rankings often do not result in a significant drift with respect to the original document [17].
[3] The cluster hypothesis was originally stated with respect to textual content of documents [22]. It could be that the document content does not change but its feature-vector representation does — e.g., due to changes of hyperlinks and anchor text.

PROPOSITION 1. *The pointwise robustness of a linear ranking function is less or equal to $\epsilon$ for any $\epsilon > 0$.*

PROOF. Let $d$ be some document. Then, $f(\vec{d} + \vec{v}) > f(\vec{d}) \Leftrightarrow \vec{w}\vec{v} > 0$. Suppose wlog that $\vec{w}[i] > 0$. We define a vector $\vec{v}$ such that $\vec{v}[j] = 0$ for all $j \neq i$ and $\vec{v}[i] = \epsilon$. Thus, we get that $\vec{w}\vec{v} > 0$ and $\rho_{point}(f; \vec{d}, q) \leq ||\vec{v}|| = \epsilon$. Consequently, $\rho_{point}(f; q) \leq \epsilon$. □

There are various ranking functions whose robustness can be substantially higher than 0. For example, in gradient boosted regression trees which are the basis of the state-of-the-art LambdaMART ranking function [27], not every change of a document's feature vector necessarily results in a retrieval score change.

#### 2.2.1 Pointwise stability.
Definition 1 refers to the worst case scenario per document $d$. That is, the minimal change to $\vec{d}$ required so as to increase $d$'s retrieval score. Indeed, as we showed above, every linear ranking function has, under this definition, robustness that approaches 0. However, linear functions differ by the extent to which a retrieval score changes with respect to a given magnitude $||\vec{v}||$ of a document change $\vec{v}$. Therefore, pointwise robustness, as defined in Definition 1, cannot be used to quantify this specific aspect of robustness, and we turn to the following definition:

DEFINITION 2 (POINTWISE STABILITY). *A ranking function $f$ is pointwise stable at level $K$ ($\in \mathfrak{R}_+$) with respect to a document change $\vec{v}$ if $\forall d$, $|f(\vec{d} + \vec{v}) - f(\vec{d})| \leq K||\vec{v}||$. The lower $K$, $f$ becomes more pointwise stable with respect to $\vec{v}$.*[4]

Contractive (Lipschitz) ranking functions $f$ are pointwise stable at level $K$ where $K$ is the Lipschitz coefficient. Various ranking functions are contractive. For example, some neural-network architectures were shown to be contractive [20][5]. Linear ranking functions are also contractive. Specifically, for $f(\vec{d}) \stackrel{def}{=} \vec{w}\vec{d} + b$ we get, using the Cauchy-Schwarz inequality:

$$|f(\vec{d} + \vec{v}) - f(\vec{d})| \leq ||\vec{w}||||\vec{v}||.$$

Hence, the Lipshitz bound $K \leq ||\vec{w}||$, and $f$ is pointwise stable at level $K \leq ||\vec{w}||$. This observation has an important implication. In linear ranking functions, such as RankSVM, the loss function used for training is often regularized by adding $\lambda||\vec{w}||$ where $\lambda$ is the parameter that controls the extent of regularization. Higher value of $\lambda$ results in decreased $||\vec{w}||$ (i.e., stronger regularization) as the goal is to minimize the objective function (loss+regularization). We thus arrive to:

COROLLARY 1. *Ceteris paribus, stronger regularization of linear ranking functions increases pointwise stability.*

We note that when using $L_2$ norms, decreasing the norm of $\vec{w}$ (i.e., increasing regularization) is intended to improve generalization and prevent overfitting; in other words, stronger regularization increases the bias of the ranking function and decreases its variance. We re-visit this point below.

The pointwise analysis treats documents independently of each other. However, ranking is determined by the relative retrieval scores of documents. Specifically, a change to a retrieval score, as large as it may be, need not necessarily affect ranking. Hence, we now turn to address the robustness of ranking functions in terms of the relative ranking of pairs of documents.

### 2.3 Pairwise Robustness

DEFINITION 3 (PAIRWISE ROBUSTNESS). *Let $d_1$ and $d_2$ be two documents such that $f(\vec{d_1}) \geq f(\vec{d_2})$. The pairwise robustness of $f$ with respect to $d_1, d_2$ and the query $q$ is: $\rho_{pair}(f; \vec{d_1}, \vec{d_2}, q) \stackrel{def}{=} \min_{\vec{v} \in \mathfrak{R}^m} ||\vec{v}||$ s.t. $f(\vec{d_2} + \vec{v}) > f(\vec{d_1})$. The pairwise robustness of $f$ with respect to the query is the expectation over document pairs: $\rho_{pair}(f; q) \stackrel{def}{=} \mathbb{E}_{d_1 \sim \mathbb{P}(\mathcal{D}), d_2 \sim \mathbb{P}(\mathcal{D} \setminus \{d: f(d) > f(d_1)\})} \rho_{pair}(f; \vec{d_1}, \vec{d_2}, q).$*

Pairwise robustness generalizes pointwise robustness: setting $d_1 = d_2$ in Definition 3 results in the pointwise robustness definition from Definition 1. Accordingly, it directly follows that:

PROPOSITION 2. *Pairwise robustness entails pointwise robustness but the reverse does not hold. That is, $\rho_{pair}(f; q) \geq \rho_{point}(f; q)$.*

In other words, the minimal document change (on average) needed to change the relative ranking of a document with respect to another document is higher than the minimal change needed to have the document's retrieval score increase. This trivial observation touches on the fundamental difference between treating documents independently and accounting for their relations.

Specifically, an important difference between pointwise and pairwise robustness is the potential dependence of the latter on distances between documents' feature vectors. For example, we now show that for a linear ranking function $f(\vec{d}) \stackrel{def}{=} \vec{w}\vec{d} + b$, the smallest change of $\vec{d_2}$ required to increase its retrieval score beyond that of $\vec{d_1}$ approaches its distance from $\vec{d_1}$. Furthermore, we show that for any linear ranking function, there exists a pair of documents' feature vectors for which the pairwise robustness of $f$ is not smaller than the distance between $\vec{d_1}$ and $\vec{d_2}$.

PROPOSITION 3. *Let $f$ be a linear ranking function. Then, $\forall d_1, d_2$ s.t. $d_1 \neq d_2$ and $f(\vec{d_1}) \geq f(\vec{d_2})$ and $\forall \epsilon > 0$, $\rho_{pair}(f; \vec{d_1}, \vec{d_2}, q) \leq ||\vec{d_1} - \vec{d_2}|| + \epsilon$. And, $\exists \vec{d_1}, \vec{d_2}$ s.t. $\rho_{pair}(f; \vec{d_1}, \vec{d_2}, q) > ||\vec{d_1} - \vec{d_2}||$.*

PROOF. Suppose wlog that $\vec{w}[i] > 0$. Fix $\epsilon > 0$ and some $d_1$ and $d_2$ ($\neq d_1$) s.t. $f(\vec{d_1}) \geq f(\vec{d_2})$. Let $\vec{v} = \vec{d_1} - \vec{d_2} + \vec{\delta}$ where $\vec{\delta}[i] = \epsilon$ and $\vec{\delta}[j] = 0$ for $j \neq i$. Then, $f(\vec{d_2} + \vec{v}) - f(\vec{d_1}) > 0$ iff $\vec{w}[i]\epsilon > 0$. Since $\rho_{pair}(f; d_1, d_2, q) \leq ||\vec{v}||$ and by the triangle inequality $||\vec{v}|| \leq ||\vec{d_1} - \vec{d_2}|| + \epsilon$, we get that $\rho_{pair}(f; \vec{d_1}, \vec{d_2}, q) \leq ||\vec{d_1} - \vec{d_2}|| + \epsilon$.

We now turn to prove the second part of the proposition. Let $\vec{d_1} \stackrel{def}{=} -\vec{w}$ and $\vec{d_2} \stackrel{def}{=} -2\vec{w}$; thus, $f(\vec{d_1}) > f(\vec{d_2})$. Furthermore, $\forall \vec{v}$: $f(\vec{d_2} + \vec{v}) - f(\vec{d_1}) > 0$ iff $\vec{v}\vec{w} > ||\vec{w}||^2$. By the Cauchy-Schwarz inequality, $\vec{v}\vec{w} \leq ||\vec{v}||||\vec{w}||$. Thus, $f(\vec{d_2} + \vec{v}) - f(\vec{d_1}) > 0 \Rightarrow ||\vec{v}|| > ||\vec{w}|| = ||\vec{d_1} - \vec{d_2}||$. Thus, $\rho_{pair}(f; \vec{d_1}, \vec{d_2}, q) > ||\vec{d_1} - \vec{d_2}||$. □

Additional question we are interested in is the connection between pointwise stability of retrieval scores and pairwise robustness[6], specifically, for linear functions: $f(\vec{d}) \stackrel{def}{=} \vec{w}\vec{d} + b$.

---

[4] Obviously, a function stable at a level $K$ is also stable at any level $> K$.
[5] Some classification architectures were shown to be contractive [20]. It can be shown that using these architectures for regression (ranking) also yields a contractive function.
[6] Recall that we showed above that the pointwise robustness of linear ranking functions approaches 0; however, their pointwise stability varies across different functions.

PROPOSITION 4. *Let $f$ be a linear ranking function. There exist $\vec{d}_1$ and $\vec{d}_2$ s.t. $\rho_{pair}(f; \vec{d}_1, \vec{d}_2, q) > K_{min}$ where $K_{min}$ is the minimal level at which $f$ is pointwise stable with respect to any $\vec{v}$.*

PROOF. Let $\vec{d}_1 = 2\vec{w}$ and $\vec{d}_2 = \vec{w}$. Note that $f(\vec{d}_1) > f(\vec{d}_2)$. For any $\vec{v}$, $f(\vec{d}_2 + \vec{v}) > f(\vec{d}_1)$ iff $\vec{w}\vec{v} > ||\vec{w}||^2$. Thus, by the Cauchy-Schwarz inequality, $||\vec{w}||^2 < \vec{w}\vec{v} \leq ||\vec{w}||||\vec{v}||$. Consequently, $\rho_{pair}(f; \vec{d}_1, \vec{d}_2, q) \geq ||\vec{v}|| > ||\vec{w}||$. As shown in Section 2.2, for linear ranking functions $|f(\vec{d} + \vec{v}) - f(\vec{d})| \leq ||\vec{w}||||\vec{v}||$ and therefore $K_{min} \leq ||\vec{w}|| < \rho_{pair}(f; d_1, d_2, q)$. □

Proposition 4 might seem counter intuitive at first glance. That is, the higher $K_{min}$, which means reduced stability, the higher $\rho_{pair}(f; \vec{d}_1, \vec{d}_2, q)$ can be — the pairwise robustness for *some* pair of documents (or more precisely, their feature vectors). However, recall that $K_{min}$ is determined with respect to all possible $\vec{v}$. Thus, it suffices that there is some $\vec{v}$ whose addition to some $\vec{d}$ increases substantially $d$'s retrieval score and hence $K_{min}$. The effect on $\rho_{pair}(f; \vec{d}_1, \vec{d}_2, q)$ can only be an increase or no change.

There are two implications of these observations. First, pointwise stability and pairwise robustness are, in general, complementary properties of a ranking function. Second, pointwise stability should mainly be used to contrast different ranking functions with respect to the same document change $\vec{v}$.

*2.3.1 Pairwise stability.* Definition 3, as was the case for Definition 1, refers to the worst case scenario: the minimal extent of a change that can be introduced to document $d_2$ which is ranked lower than $d_1$ so as to have it ranked higher than $d_1$. However, this robustness definition does not allow to compare the effect of a given fixed change on rankings induced by different rankers.

To address this task, we observe the following. Let $d_1$ and $d_2$ be two documents. Their relative ranking is determined by $sign(f(\vec{d}_2) - f(\vec{d}_1))$. Thus, to estimate whether the relative ranking of the two documents changes after $\vec{d}_2$ was changed by $\vec{v}$, we can examine

$$\Delta(\vec{d}_1, \vec{d}_2; \vec{v}) \stackrel{def}{=} |(f(\vec{d}_2 + \vec{v}) - f(\vec{d}_1)) - (f(\vec{d}_2) - f(\vec{d}_1))|.$$

The lower $\Delta(\vec{d}_1, \vec{d}_2; \vec{v})$ the higher the likelihood that the score difference after the document change has the same sign as that before the change; that is, the higher the likelihood that the relative ranking would not change. Now, $\Delta(\vec{d}_1, \vec{d}_2; \vec{v}) = |f(\vec{d}_2 + \vec{v}) - f(\vec{d}_2)|$. Thus, we are led to the following definition and consequence:

DEFINITION 4 (PAIRWISE STABILITY). *A ranking function $f$ is pairwise stable at level $K$ ($\in \Re_+$) with respect to a document change $\vec{v}$ if $\forall d_1\ d_2$, $\Delta(\vec{d}_1, \vec{d}_2; \vec{v}) \leq K||\vec{v}||$.*

And, we got that $f$ is pairwise stable at level $K$ iff it is pointwise stable at level $K$; i.e., $\forall d$, $|f(\vec{d} + \vec{v}) - f(\vec{d})| \leq K||\vec{v}||$.

Note that we essentially used $f$ to classify pairs of documents. This practice is also the basis of RankSVM [12]: a ranking function is learned by using it as a classifier upon pairs of documents. We argued that classification decisions are stable if the difference between retrieval scores after a document has changed is close to the difference before the change. A similar argument was used in recent work on estimating classifier stability with respect to a single object that has changed [20].

At the technical level, the pair-classification practice we have taken resulted in backing off from addressing both documents involved to only the one which has changed. Specifically, pairwise stability at a level $K$ is attained with respect to a given change $\vec{v}$ for *all* document pairs iff the ranking function is pointwise stable at a level $K$ with respect to $\vec{v}$; i.e., the *same* change introduced to any document is not likely to result in a rank swap of this document with another document if the change of the document's retrieval score is not large. It is therefore important to highlight an additional difference between the definitions of pairwise robustness and stability. Robustness is computed as the worst case scenario for a *given* pair of documents and is *aggregated* over all document pairs via the expectation. Stability level is a constraint imposed on *all* document pairs. Accounting for all specific pairwise relations to derive a bound on stability level is a highly difficult task.

In Section 2.2.1 we showed that for a linear ranking function $f$, $|f(\vec{d} + \vec{v}) - f(\vec{d})| \leq ||\vec{w}||||\vec{v}||$. Since $\Delta(\vec{d}_1, \vec{d}_2; \vec{v}) = |f(\vec{d}_2 + \vec{v}) - f(\vec{d}_2)|$ we get that $f$ is pairwise stable at a level $K \leq ||\vec{w}||$. Recall that the stronger the regularization of a linear function trained by minimizing loss + $\lambda||\vec{w}||$, i.e., the higher the value of $\lambda$, the lower $||\vec{w}||$. Hence, for a given small document change (i.e., small $||\vec{v}||$) we get that stronger regularization results in decreased likelihood of rank swaps. In other words:

COROLLARY 2. *Ceteris paribus, stronger regularization of linear ranking functions results in more robust rankings; that is, the rankings are less likely to change as a result of changing documents.*

Herein we use the terms "ranking robustness" and "ranking stability" interchangeably so as to refer to changes, or lack thereof, of a given ranking of a document list with respect to documents' changes; this is in contrast to the pointwise and pairwise analysis where we used the terms "robustness" and "stability" to refer to different notions. Measuring the robustness of a given ranking with respect to given documents' changes is a task we address in Section 2.5. Before delving into the details, we briefly discuss simultaneous changes of two documents. Then, in Section 2.4, we further discuss the connection between regularization and ranking robustness.

*Simultaneous change of two documents.* Heretofore, we discussed the pairwise case with respect to a change, $\vec{v}$, of one of the two documents involved. Now, suppose that document $d_1$ is changed by $\vec{v}_1$ and document $d_2$ is simultaneously changed by $\vec{v}_2$. Similarly to the case above, we can examine

$$\Delta(d_1, d_2; \vec{v}_1, \vec{v}_2) \stackrel{def}{=} |(f(\vec{d}_2 + \vec{v}_2) - f(\vec{d}_1 + \vec{v}_1)) - (f(\vec{d}_2) - f(\vec{d}_1))|.$$

We can then adapt Definition 4 (pairwise stability) as follows:

DEFINITION 5 (SIMULTANEOUS PAIRWISE STABILITY). *A ranking function $f$ is simultaneously pairwise stable at level $K$ ($\in \Re_+$) with respect to pairwise simultaneous changes $\vec{v}_1$ and $\vec{v}_2$ if $\forall d_1\ d_2$, $\Delta(\vec{d}_1, \vec{d}_2; \vec{v}_1, \vec{v}_2) \leq K(||\vec{v}_1|| + ||\vec{v}_2||)$.*

By the triangle inequality,

$$\Delta(\vec{d}_1, \vec{d}_2; \vec{v}_1, \vec{v}_2) \leq |f(\vec{d}_1 + \vec{v}_1) - f(\vec{d}_1)| + |f(\vec{d}_2 + \vec{v}_2) - f(\vec{d}_2)|.$$

Thus, if $f$ is pointwise stable with respect to each of $\vec{v}_1$ and $\vec{v}_2$ at a level $K$, it is simultaneously pairwise stable at level $K$.

For a linear ranking function $f(\vec{d}) \stackrel{def}{=} \vec{w}\vec{d} + b$ we get that:

$$\Delta(\vec{d_1}, \vec{d_2}; \vec{v_1}, \vec{v_2}) = \qquad (1)$$
$$|\vec{w}\vec{v_2} - \vec{w}\vec{v_1}| \leq ||\vec{w}||||\vec{v_2} - \vec{v_1}|| \leq ||\vec{w}||(||\vec{v_1}|| + ||\vec{v_2}||).$$

Thus, the function is simultaneously pairwise stable at level $||\vec{w}||$. Hence, we arrive again to the conclusion that stronger regularization of linear ranking functions, which results in reduced $||\vec{w}||$, yields rankings of increased robustness.

### 2.4 The Variance Conjecture

We established above the connection between stronger regularization of linear ranking functions and increased ranking robustness. Stronger regularization results in decreased variance and increased bias of the ranking function.

For example, in RankSVM [12], the width of the margin of the separating hyperplane is $\frac{2}{||\vec{w}||}$. The lower $||\vec{w}||$, as a result of stronger regularization, the wider the margin. This results in higher bias and lower variance of the learned ranking function. Indeed, larger margin means a more "robust/stable" decision surface, which resonates with the fact that ranking robustness is higher. Accordingly, we post the following variance conjecture for any ranking function:

CONJECTURE 1 (THE VARIANCE CONJECTURE). *The lower the variance of a learned ranking function the more robust the rankings induced by the function*[7].

In Section 3 we provide some empirical support to the conjecture. Specifically, we show that reducing the variance of RankSVM, which is a linear ranker, and that of LambdaMART which is a non-linear ranker, results in increased ranking robustness.

### 2.5 Measuring Ranking Robustness

One of our goals is to empirically contrast the robustness of rankings induced by different ranking functions.

Suppose that $L$ is a ranked document list retrieved using some ranking function. Suppose that some of the authors of documents in $L$ changed them (e.g., to have the documents ranked higher). After the changes, the ranking function is used to rank the documents in $L$ again; the result is a ranked list $L'$. The question is how to quantify the robustness of $L$ with respect to the changes of documents it contains given that the resultant (re-)ranked list is $L'$.

We can use any inter-ranking similarity or distance measure to quantify robustness. To compute these measures for $L$ and $L'$, we consider a document $d$ before its change, and after its change, denoted $d'$, as the same item.

Kendall's-$\tau$ *distance* (**KT** in short) is defined as the number of discordant pairs between two paired lists normalized with respect to the number of pairs of items in a list; its value is in $[0, 1]$. A discordant pair is two items whose relative ranking in one list is different than that in the other list. The higher the value, the higher the "distance" between the two lists[8].

---
[7]Low variance corresponds to improved generalization of a ranker to unseen queries. This is a different notion of robustness than that we focus on here.
[8]In contrast to Kendall's-$\tau$ distance, Kendall's-$\tau$ coefficient also considers concordant pairs and its value is in $[-1, 1]$.

KT treats equally swaps between documents at high ranks and at low ranks. However, the former have higher effect on precision-based retrieval effectiveness measures (e.g., average precision, p@k, NDCG) than the latter. Thus, we also consider the inter-list similarity measure **RBO** (rank-biased overlap) that differentiates swaps according to their ranks [26]. (We set the free parameter of RBO, $p$, to 0.7.) Additional robustness measure we are interested in is "top change" (**TC**): the value is 1 if the highest ranked document in $L$ and $L'$ is different, and 0 otherwise.

All three measures just discussed do not consider the amount of document change. However, changes in ranking due to small (potentially indiscernible) document changes are a stronger evidence for reduced robustness than those that result from large document changes. Indeed, our pointwise (Definition 2), pairwise (Definition 4) and simultaneous pairwise (Definition 5) stability definitions capture this notion. Hence, we define normalized measures of ranking robustness where normalization is with respect to the extent to which two documents have changed. We normalize KT and TC; normalizing RBO is more evolved and left for future work.

Let $d_1$ and $d_2$ be documents in $L$ that might have changed to $d_1'$ and $d_2'$ in $L'$, respectively. Inspired by Definitions 4 and 5 and Equation 1, we use the sum, difference and relative functions to quantify changes to $d_1$ and $d_2$:

$$\delta_{sum}(d_1, d_2) \stackrel{def}{=} ||\vec{d_2'} - \vec{d_2}|| + ||\vec{d_1'} - \vec{d_1}||,$$

$$\delta_{diff}(d_1, d_2) \stackrel{def}{=} \left| ||\vec{d_2'} - \vec{d_2}|| - ||\vec{d_1'} - \vec{d_1}|| \right|,$$

$$\delta_{rel}(d_1, d_2) \stackrel{def}{=} ||(\vec{d_2'} - \vec{d_2}) - (\vec{d_1'} - \vec{d_1})||.$$

To normalize KT, rather than count the number of discordant pairs of documents, for each discordant pair $(d_1, d_2)$ we use the value: $\frac{1}{\delta(d_1, d_2)+1}$. We then simply sum the values for all discordant pairs. The resulting distance measures are denoted **KT-sum**, **KT-diff** and **KT-rel**, respectively. Note that the smaller the changes of two documents which are a discordant pair, the higher the effect of the swap between them — i.e., the higher the distance and the lower the robustness. It is also important to note that the values of these measures are $\geq 0$ but are not bound from above. If we were to use the standard weighted Kendall-$\tau$ distance (cf., [19]), then we should have normalized by the sum of $\frac{1}{\delta(d_1,d_2)+1}$ over all documents pairs. Then, the upper bound would have been 1. However, while weighted Kendall-$\tau$ differentiates the effect of discordant pairs on the distance between two rankings, it does not allow to effectively compare distances between different pairs of lists. A case in point, if a list $L$ contains two documents, then the standard weighted Kendall's-$\tau$ distance between $L$ and $L'$ is equivalent to TC which does not take into account the amount of documents' changes that led to ranking changes.

To normalize TC we do the following. Suppose that $d_1$ is the highest ranked document in $L$ and $d_2'$ ($\neq d_1'$) is the highest ranked document in $L'$. We attribute the value $\frac{1}{\delta(d_1,d_2)+1}$ to this change of the highest ranked document. Thus, the more $d_1$ and $d_2$ were changed to become $d_1'$ and $d_2'$, respectively, the less weight we attribute to this change. The resultant measures are: **TC-sum**, **TC-diff** and **TC-rel**.

## 3 EMPIRICAL EXPLORATION

The goal of the exploration we present next is two-fold. First, studying the connection between the regularization of a linear ranking function and the robustness of the rankings it induces to (adversarial) documents' manipulations. According to the formal analysis presented in Section 2.3, the stronger the regularization, the more robust the rankings should be. The second goal is to empirically study Conjecture 1. That is, we explore the connection between the variance of the learned ranking function and ranking robustness using two ranking functions.

### 3.1 Experimental Setup

*3.1.1 Dataset.* To study the effect of adversarial document manipulations on ranking robustness, we used a recently published dataset that was created as a result of an on-going ranking competition [17]. The dataset is available at https://github.com/asrcdataset/.

The competition involved 31 repeated matches that lasted for 8 rounds; each match was with respect to a different query. The queries are a subset of TREC 2009-2012 topic titles that have a clear commercial intent and were more likely to stir up the competition. Students in an information retrieval course served as documents' authors. In the first round, in addition to the query itself, students were provided with an example relevant document, and were incentivized by bonus points to the course's grade to modify their documents so as to have them ranked as high as possible in the next round. Thus, the documents' modifications applied by the students can be considered as adversarial. There was no incentive, however, for producing documents relevant to the queries except for some general encouragement. Indeed, the percentage of relevant documents produced was declining throughout the competition [17]. Starting from the second round, students were presented with the ranking as well as the content of all documents submitted in the previous round in the match. To alleviate the task, documents were plain text and their length was restricted to 150 terms. Students had no prior knowledge of the ranking function and all data was anonymized. Statistics of the dataset is provided in Table 1.

The ranking function in all matches throughout the competition was based on LambdaMART [27] applied with a subset of the content-based features from Microsoft's learning-to-rank datasets[9]. In addition, several external modifications were made to ensure high ranking quality. First, each document in the dataset was labeled as valid, keyword stuffed or spam by five Figure Eight (https://www.figure-eight.com/) annotators. (All annotations are available as part of the dataset.) Documents with at least four non-valid labels were penalized in the ranking. Second, documents similar to the example relevant document were demoted in the ranking. Third, documents were penalized in rankings if their content was duplicated from other documents in the previous round.

*3.1.2 Ranking Functions.* The exploration we present uses the documents created in the ranking competition just described. Rather than using the original ranking induced for each query in each round of the competition, we re-rank the documents using the ranking functions we study. One advantage of this practice is that we refrain from infusing noise caused by external interventions as those describe above (e.g., penalizing documents whose content was duplicated), and we focus on the ranking functions of interest. Thus, while the ranking functions we study were not exactly those used in the competition, the manipulations of documents were adversarial as the only incentive was to have them ranked high, and the basic ranking function employed in the competition, LambdaMART, is highly effective. In other words, we measure ranking robustness by examining adversarial manipulations of documents introduced in response to a strong ranker.

Table 1: The dataset used for experiments [17].

| | | | |
|---|---|---|---|
| Queries | 31 | Queries per student | 3 |
| Rounds | 8 | Students per query | 5 – 6 |
| Students | 52 | Unique documents | 897 |
| Documents | 1279 | Relevant documents | 1113 |

We learned two families of ranking functions: RankSVM and LambdaMART. As was the case for the LambdaMART ranker used in the ranking competition, the ClueWeb09 Category B collection, henceforth **ClueWeb**, was used to train all ranking functions[10]. Titles of topics 1-200 from TREC 2009-2012 served for queries. We applied Krovetz stemming and removed stopwords on the INQUERY list from queries only. The Indri toolkit[11] was used for experiments. To learn the ranking functions, 75% randomly sampled queries served for training and the remaining 25% for validation of hyper-parameter values. Once hyper-parameter values were set, either using the procedure just described or to some predefined values as detailed below, all the queries were used to train the final ranking functions. These were then applied to rank documents in each round of the competition for each query. When we present evaluation for ClueWeb, we report the results of using five-fold cross validation. NDCG@20 served as the optimization criterion in all cases.

The ranking functions we use are trained with, and applied on, document-query feature vectors composed of 26 content-based features; 25 of these features are those used in the ranking competition [17]. We generated an additional query-independent document quality feature as follows. Instead of directly penalizing non-valid documents as was the case in the competition, we used these annotations to simulate Waterloo's spam classification scores [3]. Specifically, we used $100 - 20(k + s)$, where $k$ and $s$ are the number of keyword-stuffed and spam labels a document received, respectively; the scores for the example relevant documents were set to 100. Since $k + s \leq 5$, the simulated scores are in $[0, 100]$ as is the case for Waterloo's scores[12]. Thus, while the learned ranking function is based on Waterloo's scores, for the competition dataset it is applied with the human-annotation-based score we created which serves as a document quality measure.

*3.1.3 Outline of Experiments.* The goal of the regularization-based experiments we report is to study the connection between the regularization of a linear RankSVM and LambdaMART and the robustness and effectiveness of rankings they induce. An additional experiment is intended to contrast the ranking robustness of the two rankers. Hence, we now turn to describe the regularization approaches we have employed for the rankers.

---
[9]https://tinyurl.com/rmslr
[10]The document collection created in the competition is too small to effectively learn ranking functions.
[11]www.lemurproject.org/indri
[12]Waterloo's classifier scores represent the percentage of documents in a collection that are "spammier" than the given document.

*Regularization of RankSVM.* The objective function of RankSVM can be expressed using hinge loss:

$$\min_{\vec{w}} L(\vec{w}; c) \stackrel{def}{=} \frac{1}{2}||\vec{w}||^2 + c \sum_{i,j,t} \max(0, 1 - \vec{w}(d_i^{\vec{q}_t} - d_j^{\vec{q}_t})l_{d_i^{q_t}, d_j^{q_t}}^{q_t});$$

$c(> 0)$ is a regularization parameter; $d_i^{q_t}$ and $d_j^{q_t}$ are two documents retrieved for query $q_t$; $l_{d_i^{q_t}, d_j^{q_t}}^{q_t} = 1$ if the relevance label for $q_t$ of $d_i^{q_t}$ is higher than that of $d_j^{q_t}$ and $-1$ otherwise. The objective function can be re-written as follows [16]:

$$\min_{\vec{w}} L(\vec{w}; \lambda) = \frac{\lambda}{2}||\vec{w}||^2 + \frac{1}{n}\sum_{i,j,t} \max(0, 1 - \vec{w}(d_i^{\vec{q}_t} - d_j^{\vec{q}_t})l_{d_i^{q_t}, d_j^{q_t}}^{q_t});$$

$n$ is the number of document pairs across queries used for training; $c = \frac{1}{n\lambda}$. Thus, the lower the value of $c$, the higher the value of $\lambda$, and hence the lower $||\vec{w}||$ — i.e., the regularization is stronger. Indeed, lower $c$ corresponds to higher margin of the separating hyperplane ($\frac{2}{||\vec{w}||}$), and therefore to reduced variance and increased bias.

One of our goals is to study the correlation between $||\vec{w}||$ and ranking effectiveness and robustness. To that end, we vary the value of $c$ for learning $\vec{w}$, and study the impact on induced rankings. We set $c$ in these experiments to 189 different values in $[0, 10000]$[13]. We used $SVM^{rank}$[14] to train RankSVM; except for the value of $c$, all other parameters were set to default values. We found high positive correlation between the value of $c$ and the resultant $||\vec{w}||$. Herein, correlations are measured using Spearman coefficient, which is an estimate of the monotonic relationship between variables, Pearson correlation, which is an estimate for linear relationship, and Kendall's-$\tau$. The values of all three measures are in $[-1, 1]$: $-1$ and $1$ indicate perfect negative and positive correlation, respectively. The statistical significance of correlations (with respect to 0) is determined at a 95% confidence level. The Spearman, Pearson and Kendall correlations between $c$ and $||\vec{w}||$ are 0.951, 0.448 and 0.867, respectively; all the correlations are statistically significant[15].

As noted above, in another experiment we performed, we compared the robustness of rankings induced by RankSVM and LambdaMART. Here we set $c$ to values in $\{0.001, 0.01, 0.1\}$ using the 75-25 split for queries over ClueWeb described above. Recall that NDCG@20 was the optimization criterion. The ranking competition did not have enough rounds so as to optimize directly for robustness and then test it. Hence, we optimized for ranking effectiveness and evaluated the effect on robustness. We hasten to point out, however, that these values of $c$ resulted in high ranking robustness. We describe below how we measured robustness.

*Regularization of LambdaMART.* LambdaMART uses gradient boosted regression trees. The higher the number of trees and leaves, the lower the regularization, and the higher the variance.

We used RankLib's[16] implementation of LambdaMART. To study the connection between the regularization of LambdaMART and the effectiveness and robustness of rankings it induces, we set (#leaves, #trees) to values in $\{(5, 150), (10, 160), (15, 170), \ldots, (150, 440)\}$.

To compare the ranking robustness of LambdaMART with that of RankSVM, we set the number of leaves and trees to values in $\{250, 500\}$ and $\{25, 50\}$, respectively, using the 75-25 split mentioned above. These values resulted in high ranking robustness. We now turn to describe how ranking robustness was evaluated.

*3.1.4 Evaluating Ranking Robustness.* Suppose that at round $i$ of the competition for query $q$ the set of documents created by the students is $S$. Now, we rank $S$ using a ranking function to produce a ranked list $L_i$. Then, we consider the set of documents created by the same students in round $i + 1$ of the competition for query $q$. Ranking this set yields the list $L_{i+1}$. We can then apply upon $L_i$ and $L_{i+1}$ each of the ranking-robustness measures we described in Section 2.5. For each query, we average these values over pairs of rounds $i, i + 1$; we then average the resultant values over queries. Thus, for each ranking function and robustness measure we get a single ranking robustness value.

Accordingly, we can measure the correlation between the regularization strength of RankSVM (as manifested in different values of $||\vec{w}||$) and ranking robustness; and, between LambdaMART's regularization strength (with (#leaves, #trees) serving as a proxy) and its ranking robustness. As at the above, we use Spearman, Pearson and Kendall's-$\tau$ to measure correlations. Then, to contrast the ranking robustness of RankSVM and LambdaMART, we compare the values of the robustness measures. Statistically significant differences of retrieval performance and of values assigned by robustness measures are determined using the two-tailed paired t-test at a 95% confidence level. Recall that the robustness measure value for each query is the average over pairs of consecutive rounds. These values are compared for statistical significance over queries.

*3.1.5 Evaluating Ranking Effectiveness.* All documents in the competition were judged for relevance via Figure Eight by five annotators [17]. The relevance judgments are available as part of the dataset. Here, we consider a document relevant if it was labeled as such by at least three annotators. To estimate retrieval effectiveness we use MAP (mean average precision) and NDCG (normalized discounted cumulative gain) at cutoff 5; this is the number of students that participated in a competition per query[17].

We also evaluate retrieval effectiveness over ClueWeb. In this case, MAP was computed for the top 1000 retrieved documents and NDCG for the top 20; the latter also served as the optimization metric for training LambdaMART.

## 3.2 Experimental Results

*3.2.1 Ranking Robustness of RankSVM.* In Figure 1 we present the values of the (unnormalized) robustness measures as a function of $||\vec{w}||$ in RankSVM. Recall that high values of KT and TC attest to decreased ranking robustness, as these are measures of inter-list distance, and high values of RBO attest to increased robustness as this is an inter-list similarity measure. We see that increasing $||\vec{w}||$, which means weaker regularization, results in decreased ranking robustness when measured using KT and RBO. (The differences with TC are discussed below.) This finding is aligned with our formal analysis of linear ranking functions in Section 2.

---

[13]Random sampling of the value of $c$ results in a very short range of values of $||\vec{w}||$. Hence, we sampled from $[10^x, 9 \cdot 10^x]$ with steps of size $10^x$ for $x \in \{-4, -3, \ldots, 2\}$. We sampled from $[1000, 10000]$ with steps of 1000. We further sampled 60 values uniformly from $[0, 10000]$; in addition, we used values in $\{1, 40, 80, 120, 160, \ldots, 1000\}$ and $\{1, 45, 90, 135, 180, \ldots, 1125\}$.
[14]https://www.cs.cornell.edu/people/tj/svm_light/svm_rank.html
[15]We consider only values of $c$ for which $||\vec{w}|| < 15$. See Section 3.2.1 for details.

[16]https://tinyurl.com/ranklib
[17]To generate graded relevance judgments we followed Raifer at al. [17]: documents labeled relevant by 3, 4 and 5 annotators were considered marginally relevant, fairly relevant, and highly relevant, respectively; documents with at most 2 relevant labels were considered non-relevant.

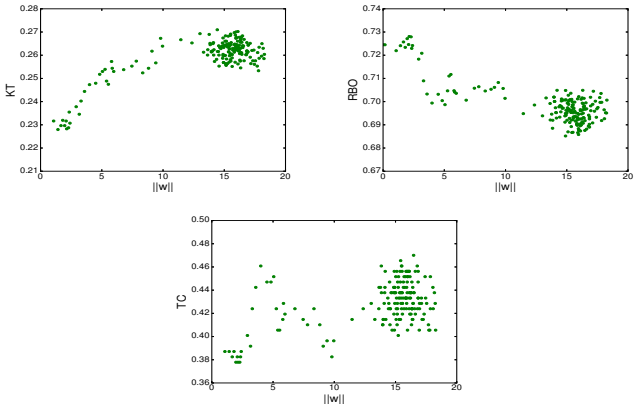

Figure 1: Robustness per $||\vec{w}||$ in RankSVM.

Table 2: The correlation between $||\vec{w}||$ and robustness measures' values for RankSVM. 'S', 'P' and 'K' stand for Spearman, Pearson and Kendall, respectively. '★': the correlation is statistically significant.

|   | KT | KT-sum | KT-diff | KT-rel | RBO | TC | TC-sum | TC-diff | TC-rel |
|---|---|---|---|---|---|---|---|---|---|
| S | .748★ | .745★ | .678★ | .737★ | −.756★ | .528★ | .574★ | .475★ | .591★ |
| P | .889★ | .872★ | .850★ | .875★ | −.841★ | .567★ | .632★ | .483★ | .656★ |
| K | .574★ | .560★ | .509★ | .562★ | −.558★ | .374★ | .413★ | .336★ | .427★ |

To compute correlations between ranking robustness measures and $||\vec{w}||$, we first observe the following in Figure 1. Our sampling of values of $c$, the regularization parameter in RankSVM, resulted in over-sampling of $||\vec{w}||$ values somewhat larger than $||\vec{w}|| = 15$. This can substantially affect correlation values. Hence, for the correlation analysis to follow, and for studying RankSVM's ranking effectiveness as a function of $||\vec{w}||$, we do not consider RankSVM functions with $||\vec{w}|| \geq 15$. The resulting sample contains 71 RankSVM functions. We hasten to point out that this sampling did not change the polarity of the correlations we found, nor their statistical significance, but only the actual correlation numbers which increased[18].

Table 2 presents the correlation between $||\vec{w}||$ and the values of all ranking-robustness measures. As can be seen, there are high, statistically significant, correlations between $||\vec{w}||$ and the values of all robustness measures for all three correlation metrics (Spearman, Pearson and Kendall)[19]. Thus, the findings in Table 2 provide support to our formal analysis with respect to linear ranking functions

[18] For example, without removing models with $||\vec{w}|| \geq 15$, the Spearman correlation for KT, RBO and TC is: .33, −.466 and .284, respectively; Pearson correlation is .806, −.780, and .503, respectively; Kendall's-τ numbers are: .246, −.343, and .217, respectively. All correlations are statistically significant.
[19] Recall that RBO is an inter-list similarity measure, in contrast to the other two; hence increasing values of RBO attest to increased robustness.

Table 3: The correlation between #leaves and #trees in LambdaMART and the values of the robustness measures. 'S', 'P' and 'K' stand for Spearman, Pearson and Kendall, respectively. '★': the correlation is statistically significant.

|   | KT | KT-sum | KT-diff | KT-rel | RBO | TC | TC-sum | TC-diff | TC-rel |
|---|---|---|---|---|---|---|---|---|---|
| S | .725★ | .733★ | .754★ | .727★ | −.600★ | .506★ | .521★ | .507★ | .547★ |
| P | .674★ | .714★ | .728★ | .732★ | −.615★ | .525★ | .530★ | .568★ | .577★ |
| K | .549★ | .568★ | .609★ | .563★ | −.453★ | .366★ | .366★ | .379★ | .398★ |

— RankSVM in this case: stronger regularization (decreased $||\vec{w}||$) results in increased ranking robustness. This also provides support to our variance conjecture (Conjecture 1): increased $||\vec{w}||$ means higher variance; thus, higher variance is indeed correlated with decreased ranking robustness for RankSVM.

In comparing the normalized robustness measures in Table 2 (X-sum, X-diff and X-rel measures) with their unnormalized counterparts we see the following. For KT, the normalized versions yield minor to moderate decrease of correlation with $||\vec{w}||$, although the correlations remain statistically significant and high. Similarly, TC-diff yields lower correlation than TC. However, TC-sum and TC-rel yield higher correlations than TC.

Table 2 also shows that in almost all cases, KT-rel yields higher correlation than KT-sum and KT-diff, and TC-rel yields higher correlation than TC-sum and TC-diff. Recall from Section 2.5 that the X-rel measures use for normalization the norm of the difference between the vector change of one document and that of the other. This norm is $||\vec{v}_2 - \vec{v}_1||$ used in the upper bound for simultaneous pairwise stability of a linear ranking function in Equation 1.

Finally, Table 2 shows that KT and its normalized variants, and RBO, yield higher correlation than TC and its variants. This is not a surprise as KT and RBO consider changes in the entire document list, while TC only considers changes at the highest rank; thus, TC is a less robust measure of ranking robustness. (These findings are aligned with the patterns observed in Figure 1.) We also see that RBO yields lower Pearson and Kendall correlation than KT. RBO attributes more weight to swaps of documents at high ranks and KT does not. However, the document lists are short (composed of 5 documents); hence, the differences in correlations potentially should not be attributed to different weighting of different ranks.

*3.2.2 Ranking Robustness of LambdaMART.* Table 3 shows the correlations between the number of leaves (#leaves) and trees (#trees) used to train LambdaMART, which were increased simultaneously, and the resulting values of ranking robustness. This increase corresponds to decreased regularization; i.e., higher variance. Indeed, Table 3 shows that according to all three correlation metrics and for all three ranking-robustness measures, robustness decreases with decreased regularization. This finding provides further support to our variance conjecture (Conjecture 1): ranking functions with higher variance yield rankings of decreased robustness. Thus, we attained support for the conjecture for two different rankers: the first is linear (RankSVM) and the second is not (LambdaMART).

Table 3 also shows that KT yields higher correlation than RBO which in turn yields higher correlation than TC. These findings are aligned with those presented above for RankSVM. We also see that the normalized versions of KT and TC (KT-X and TC-X) yield increased correlations with respect to the unnormalized versions (KT and TC). This finding implies here to the importance of accounting not only for the change of ranking, but also for the change in documents that drove the change in ranking when measuring ranking robustness. Finally, we see that for TC, the normalized version TC-rel yields the highest correlations which is aligned with the findings above for both TC and KT. However, for KT here, KT-rel does not dominate KT-sum and KT-diff.

Table 4: Comparison of RankSVM and LambdaMART in terms of ranking robustness and effectiveness. The numbers in parentheses indicate the percentage of queries for which the robustness number attained for one ranking function is higher than that attained for the other. Percentages do not necessarily sum to 100 due to ties and rounding. '★' marks statistically significant differences with LambdaMART. Bonferroni correction was applied for robustness comparisons.

|         | LambdaMART     | RankSVM         |
|---------|----------------|-----------------|
| KT      | 0.343 (54.8)   | 0.264★ (27.2)   |
| TC      | 0.599 (30.9)   | 0.401★ (11.1)   |
| RBO     | 0.613 (26.3)   | 0.703★ (60.8)   |
| TC-diff | 0.361 (53.5)   | 0.219★ (28.6)   |
| TC-rel  | 0.246 (50.7)   | 0.153★ (31.3)   |
| TC-sum  | 0.213 (51.6)   | 0.133★ (30.4)   |
| KT-diff | 2.071 (62.7)   | 1.505★ (31.8)   |
| KT-rel  | 1.399 (63.6)   | 1.008★ (30.9)   |
| KT-sum  | 1.204 (60.8)   | 0.874★ (33.6)   |
| MAP     | 0.825 (16.5)   | 0.846  (22.7)   |
| NDCG    | 0.837 (26.7)   | 0.854  (37.5)   |

#### 3.2.3 Comparing RankSVM and LambdaMART.
In Table 4 we contrast the robustness and effectiveness of rankings induced by RankSVM and LambdaMART over the ranking-competition dataset. Recall that RBO measures the similarity between two lists, hence higher values attest to higher robustness. In contrast, KT and its normalized variants as well as TC and its normalized variants measure the differences between two lists; hence, lower values attest to higher robustness. We see that the ranked lists induced by RankSVM are (on average) more robust than those induced using LambdaMART according to all considered ranking-robustness measures; all the differences are statistically significant. Furthermore, the robustness of lists induced by RankSVM surpasses that of lists induced by LambdaMART for the majority of queries. (Refer to the numbers in parentheses.) These findings provide additional support to our variance conjecture. Since RankSVM is linear and LambdaMART is not, the variance of RankSVM is in general lower, and we saw that its ranking robustness is higher.

In the last two lines of Table 4 we compare the retrieval effectiveness of RankSVM and LambdaMART. Although the differences are not statistically significant, we see that lists induced by RankSVM are not only more robust but also somewhat more effective compared to LambdaMART on the competition dataset. The high MAP and NDCG values can be attributed to the fact that most documents generated by the students were judged to be relevant. (See Table 1.)

Table 5: Correlation between regularization of RankSVM ($||\vec{w}||$) and that of LambdaMART (#leaves and #trees) and retrieval effectiveness. Competition: Raifer et al.'s dataset [17]. '★': the correlation is statistically significant.

|            |             |      | Spearman | Pearson | Kendall |
|------------|-------------|------|----------|---------|---------|
| RankSVM    | Competition | MAP  | −.777★   | −.918★  | −.559★  |
|            |             | NDCG | −.787★   | −.897★  | −.564★  |
|            | ClueWeb     | MAP  | .220★    | .549★   | .079    |
|            |             | NDCG | .259★    | .636★   | .141    |
| LambdaMART | Competition | MAP  | −.732★   | −.732★  | −.549★  |
|            |             | NDCG | −.726★   | −.710★  | −.545★  |
|            | ClueWeb     | MAP  | −.849★   | −.871★  | −.692★  |
|            |             | NDCG | −.804★   | −.812★  | −.635★  |

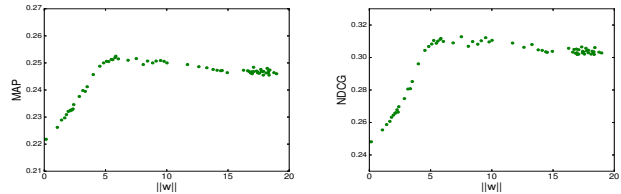

Figure 2: The connection between $||\vec{w}||$ and retrieval effectiveness (MAP and NDCG) on ClueWeb of RankSVM.

#### 3.2.4 Regularization and Retrieval Effectiveness.
Table 5 presents the correlation between retrieval effectiveness (MAP and NDCG) and regularization for RankSVM ($||\vec{w}||$) and LambdaMART (#leaves and #trees). As most documents in the ranking-competition dataset are relevant, we also analyze retrieval performance for ClueWeb that has a much higher percentage of non-relevant documents.

Table 5 shows that except for the case of RankSVM over ClueWeb, there is high statistically significant positive correlation between increased regularization and retrieval effectiveness. (Decreasing $||\vec{w}||$ and #leaves, #trees amounts to increased regularization.) Figure 2, which presents the connection between MAP/NDCG and $||\vec{w}||$ of RankSVM on ClueWeb, shows a positive trend for $||\vec{w}|| \leq 5$ and a negative trend for $||\vec{w}|| \geq 5$. Substantially increasing (decreasing) $||\vec{w}||$ can result in overfitting (underfitting). Thus, the correlations for RankSVM in Table 5 seem to be quite affected by a long range of underfitting which results in decreased retrieval performance; increasing $||\vec{w}||$ then reduces underfitting and improves performance until the ranker overfits. In Figure 3 we see that, in general, decreased regularization of LambdaMART (i.e., higher number of leaves and trees) results in decreased MAP performance over ClueWeb. This finding is aligned with those that emerge based on Table 5.

## 4 RELATED WORK

There are notions of robustness of ad hoc retrieval functions different than that we examine here: the performance variance over queries [23] and the relative performance over queries with respect to another ranking function [24]. We study changes to rankings that result from documents' (adversarial) manipulations.

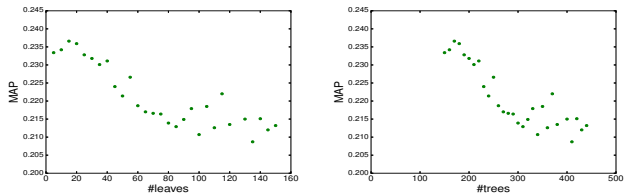

Figure 3: The connection between #leaves and #trees used in LambdaMART and its MAP effectiveness over ClueWeb.

The changes of a ranked document list that result from synthetic random noise introduced to documents were used to predict query performance [28]. We address a different setting — adversarial document changes intended to promote documents in rankings — and present a different analysis.

Much of the work on adversarial information retrieval focuses on identifying and addressing different types of spam (e.g., content-based and hyperlink-based) [1, 2]. In contrast to our work, there was no formal and empirical analysis of ranking robustness with respect to (adversarial) documents' manipulations.

Recent work analyzes the strategies employed by documents' authors in ranking competitions [17]. In contrast, we analyze the robustness of ranking functions. We use the datasets of ranking competitions organized in this work [17] in our empirical analysis.

The probability ranking principle [18] was shown to be sub-optimal in competitive retrieval settings, where authors manipulate their documents so as to have them highly ranked. However, the robustness of ranking functions was not studied as in our work.

There is a growing body of work on adversarial and robust classification; e.g., [4, 6–10, 14, 20, 21, 25]. The focus is on improving classifier's robustness to adversarial (often, minuscule) manipulations of objects and their feature values. In contrast to our work, the robustness of document ranking functions was not studied; specifically, the pairwise robustness notions we analyze, which are a core aspect of ranking robustness, were not studied.

The connection between neural network regularization and the stability of classification decisions has recently been demonstrated [20]. We demonstrate the connection between regularization of linear ranking functions and stability of retrieval scores, and more importantly, ranking robustness.

## 5 CONCLUSIONS AND FUTURE WORK

We presented a formal and empirical analysis of the robustness of rankings induced by feature-based relevance-ranking functions to (adversarial) manipulations of documents. We formally showed that increased regularization of linear ranking functions results in increased ranking robustness. Accordingly, we conjectured that increased variance of any learned ranking function results in decreased ranking robustness. We provided empirical support to our formal findings and the conjecture by analyzing ranking competitions where authors introduced adversarial changes to documents.

We plan to further study and improve the robustness of non-linear learning-to-rank functions. We also intend to extend the robustness analysis to sets of queries; e.g., those representing the same information needs.

**Acknowledgments** We thank the reviewers for their comments. This work was supported by funding from the European Research Council (ERC) under the European Union's Horizon 2020 research and innovation programme (grant agreement 740435).